\documentclass[journal]{IEEEtran}
\usepackage{makecell}
\usepackage{cite}
\usepackage{graphicx}
\usepackage{amsmath}
\usepackage{amsfonts}
\usepackage{multirow}
\usepackage{subcaption}
\usepackage{color}
\usepackage{booktabs}
\usepackage[flushleft]{threeparttable}
\usepackage{hyperref}
\usepackage{url}

\begin{document}
	%
	\title{A Novel Combined Data-Driven Approach for Electricity Theft Detection}
	%
	%
	%
	
		\author{Kedi~Zheng,~\IEEEmembership{Student~Member,~IEEE,}
			~Qixin~Chen,~\IEEEmembership{Senior~Member,~IEEE,}
			~Yi~Wang,~\IEEEmembership{Student~Member,~IEEE,}
			~Chongqing~Kang,~\IEEEmembership{Fellow,~IEEE,}
			~and~Qing~Xia,~\IEEEmembership{Senior~Member,~IEEE}
			\thanks{Manuscript received July 18, 2018; revised September 10, 2018; accepted
				September 28, 2018. This work was supported by National Key R\&D Program of China (No. 2016YFB0900100). Paper No. TII-18-1861. (\textit{Corresponding Author: Qixin Chen})}
			\thanks{The authors are with the State Key Lab of Power Systems, the Department of Electrical Engineering, Tsinghua University, Beijing, 100084 China. (E-mail: qxchen@tsinghua.edu.cn)}
			\thanks{Digital Object Identifier \href{https://doi.org/10.1109/TII.2018.2873814}{10.1109/TII.2018.2873814}}
		}
	
	%
	%
	
	\markboth{IEEE TRANSACTIONS ON INDUSTRIAL INFORMATICS, ACCEPTED FOR PUBLICATION}%
	{Shell \MakeLowercase{\textit{et al.}}: Bare Demo of IEEEtran.cls for IEEE Journals}
	%
	
	
	\maketitle

	\IEEEpubid{\begin{minipage}{\textwidth}\ \\[12pt] \centering
			© 2018 IEEE.  Personal use of this material is permitted.  Permission from IEEE must be obtained for all other uses, in any current or future media, including reprinting/republishing this material for advertising or promotional purposes, creating new collective works, for resale or redistribution to servers or lists, or reuse of any copyrighted component of this work in other works.
	\end{minipage}}

	\IEEEpubidadjcol

	\begin{abstract}
		The two-way flow of information and energy is an important feature of the Energy Internet. Data analytics is a powerful tool in the information flow that aims to solve practical problems using data mining techniques. As the problem of electricity thefts via tampering with smart meters continues to increase, the abnormal behaviors of thefts become more diversified and more difficult to detect. Thus, a data analytics method for detecting various types of electricity thefts is required. However, the existing methods either require a labeled dataset or additional system information which is difficult to obtain in reality or have poor detection accuracy. In this paper, we combine two novel data mining techniques to solve the problem. One technique is the Maximum Information Coefficient (MIC), which can find the correlations between the non-technical loss (NTL) and a certain electricity behavior of the consumer. MIC can be used to precisely detect thefts that appear normal in shapes. The other technique is the clustering technique by fast search and find of density peaks (CFSFDP). CFSFDP finds the abnormal users among thousands of load profiles, making it quite suitable for detecting electricity thefts with arbitrary shapes. Next, a framework for combining the advantages of the two techniques is proposed. Numerical experiments on the Irish smart meter dataset are conducted to show the good performance of the combined method. 
	\end{abstract}
	
	
	%
	\IEEEpeerreviewmaketitle

		\section*{Nomenclature}
		\addcontentsline{toc}{section}{Nomenclature}
		
		\noindent \textit{Sets}
		\begin{IEEEdescription}
			\item[$\mathcal{A}$] Set of all users in an area.
			\item[$\mathcal{B}$] Set of benign users in the area.
			\item[$\mathcal{F}$] Set of fraudulent users in the area.
			\item[$\mathcal{D}$] Set of ordered data pairs.
		\end{IEEEdescription}
		
		\noindent \textit{Indices}
		\begin{IEEEdescription}[\IEEEusemathlabelsep\IEEEsetlabelwidth{$p,q$}]
			\item[$t$] Time interval index. 
			\item[$i$] Electricity user index.
			\item[$p,q$] Load profile indices.
			\item[$j$] Day index.
		\end{IEEEdescription}
		
		\noindent \textit{Variables and Parameters}
		\begin{IEEEdescription}
			\item[$x_{i,t}$] Ground truth consumption for user $i$ at time $t$. 
			\item[$\tilde{x}_{i,t}$] Recorded consumption for user $i$ at time $t$.
			\item[$\textbf{x}_i$] Ground truth load profile for user $i$.
			\item[$\tilde{\textbf{x}}_i$] Recorded load profile for user $i$.
			\item[$\textbf{u}_{i,j}$] Normalized load profile for user $ i $ at day $ j $.
			\item[$E_t$] Ground truth consumption of the area at time $t$.
			\item[$e_t$] Non-technical loss of the area at time $t$.
			\item[$\textbf{e}$] Non-technical loss series of the area.
			\item[$G$] A grid used to seperate $ \mathcal{D} $ into discrete values.
			\item[$a,b$] Number of bins in x-axis and y-axis of $ G $.
			\item[$\textbf{M}$] Characteristic matrix. 
			\item[$\rho_p$] Local density of load profile $ p $.
			\item[$\delta_p$] Minimal distance of load profile $ p $ from other load profiles.
			\item[$d_{p,q}$] Distance between load profiles $ p $ and $ q $.
			\item[$d_c$] Cut-off distance for CFSFDP.
			\item[$\zeta_p$] Density abnormality for load profiles $ p $.
		\end{IEEEdescription}
		
		\noindent \textit{Functions}
		\begin{IEEEdescription}[\IEEEusemathlabelsep\IEEEsetlabelwidth{$Corr(\cdot,\cdot)$}]
			\item[$|\cdot|$] Size of a set.
			\item[$Corr(\cdot,\cdot)$] Correlation measurement for two vectors. 
			\item[$I(\cdot)$] Mutual Information of data.
			\item[$B(\cdot)$] Upper bound function for the scale of $ G $.
			\item[$MIC(\cdot)$] Maximum Information Coefficient of data.
			\item[$\chi(\cdot)$] Kernel function used in CFSFDP.
		\end{IEEEdescription}
	
	\section{Introduction}
	%
	%
	%
	%
	\IEEEpubidadjcol
	
	\IEEEPARstart{T}{he Energy} Internet, which is proposed as the next step in the evolution of Smart Grid~\cite{zhang2017distributed}, has the important feature of bi-directional energy and information flow. The advanced metering infrastructure (AMI) is the basis of the information flow in the Energy Internet. With the deployment of smart meters, AMI now provides power utilities with massive amounts of electricity consumption data at a higher frequency, thus enabling precise user behavior modeling~\cite{wang2016clustering}, load forecasting~\cite{wang2018review}, load estimation~\cite{sun2018probabilistic}, and demand response~\cite{sijie2017demand}. However, making the information flow of Energy Internet secure has proved to be a challenging issue due to the unique characteristics of AMI. Fraudulent users can tamper with the smart meter data using digital tools or cyber attacks. Thus, the form of electricity thefts in Energy Internet is very different from the thefts in the past, which relies mostly on physically bypassing or destructing mechanical meters~\cite{jiang2014energy}.

	Cases of organized energy theft spreading tampering tools and methods against smart meters that caused severe loss of power utilities were reported by the U.S. Federal Bureau of Investigation~\cite{fbi2012cyber} and Fujian Daily~\cite{fujian2013first} in China. In total, the non-technical loss (NTL) due to consumer fraud in the electrical grid in the U.S. was estimated to be \$6 billion/year ~\cite{mcdaniel2009security}. Because the traditional detection methods of sending technical staff or Video Surveillance are quite time-consuming and labor-intensive, electricity theft detection methods that take the advantage of Energy Internet's information flow are urgently needed to solve the problem of the "Billion-Dollar Bug". 
	
	The existing non-hardware electricity theft detection methods can be classified into three categories: artificial intelligence-based (AI-based), state-based, and game theory-based~\cite{jokar2016electricity}. The AI-based methods use machine learning techniques, such as classification and clustering to analyze the load profiles of consumers to find the abnormal users because the consumption patterns of fraudulent users are believed to differ from those of benign users. Classification methods~\cite{nizar2008power,zheng2018wide,ahmad2017review} usually require the labeled dataset to train the classifier, whereas clustering methods~\cite{junior2016unsupervised,zanetti2017tunable,sun2017cvine} are unsupervised and can be applied to an unlabeled dataset. The state-based methods~\cite{neto2013probabilistic,leite2017detecting} use additional measurements, such as power, voltage, and current in the distribution network to detect electricity thefts. Because fraudulent users are incapable of tampering with the network measurements, so conflicts will arise between the system states and smart meter records. Although high detection accuracy can be achieved, these methods require the network topology and additional meters. The game theory-based methods~\cite{cardenas2012game,amin2015game} assume that there is a game between fraudulent users and power utilities and that different distributions of fraudulent users' and benign users' consumption can be derived from the game equilibrium. Detection can be conducted according to the difference between the distributions. Because the game theory-based methods focus on theoretical analysis with strong assumptions, they are beyond the scope of this paper. 
	
	A brief review of the existing state- and AI-based electricity theft detection methods in the literature is presented here. 
	
	The physical model of a power network indicates that the system variables should satisfy certain mathematical equations, which derives the consistency of the variables. The state-based methods utilize the fact that tampering with smart meter data will certainly create inconsistencies between system variables including power, voltage and current. In~\cite{nikovski2014method}, a linear regression method is used to estimate the resistance of distribution lines from active power and current measurements; next, the NTL of each line is calculated according to the estimated resistance value to find the electricity theft. A state-estimation-based approach for distribution transformer load estimation is exploited to detect
	meter tampering in~\cite{huang2013non}. The variance of measurements
	and estimated values is analyzed to create a suspect list of customers with metering
	problems. Neto \textit{et al.} proposed a probabilistic methodology for NTL estimation in the distribution network~\cite{neto2013probabilistic}. The technical loss sensitivity in relation to the load variation is derived and the probabilistic distributions of total loss and technical loss are calculated. In their methodology, if the two distributions have big differences, then the NTL is indicated. Han \textit{et al.} proposed a fast NTL fraud detection (FNFD) scheme in~\cite{han2016fnfd}, where the NTL is calculated from observer meters and the Recursive Least Square (RLS) algorithm is used to find the correlation between smart meter data and the NTL. FNFD can catch proportional electricity thieves who steal energy at a fixed proportion. In~\cite{he2017real}, a deep-learning-based real-time mechanism for detecting electricity thefts was proposed. In this mechanism, the state vector estimator (SVE) calculates the attack vector and the state vector from real-time measurements and the power system topology, and an identification scheme based on a deep belief network helps the SVE finds the false data injection (FDI). Xiao \textit{et al.} proposed an algorithm for regional and individual electricity theft detection using random matrix theory (RMT) in~\cite{xiao2017electricity}. A pattern signal is constructed from real time power and voltage measurements as an indicator for NTL. Most of the state-based methods rely on the real-time acquisition of the system topology and additional physical measurements, which is sometimes unattainable. 
	
	In almost all occasions, the tampered load profiles differ from the original ones. The AI-based methods attempt to find the abnormal patterns among all load profiles of the consumers. Nizar \textit{et al.} applied the Extreme Learning Machine (ELM) for electricity theft detection \cite{nizar2008power}. The ELM-based approach extracts patterns of customer behavior from historical kWh consumption data and detects abnormal behaviors. In~\cite{jokar2016electricity}, a multi-class support vector machine (SVM) was trained to detect whether a new sample of load profiles is normal or malicious. The problem of imbalanced training dataset is addressed and solved by generating a synthetic dataset. In~\cite{zheng2018wide}, a Wide \& Deep Convolutional Neural Networks (CNN) model was developed and applied to analyze the electricity thefts in Smart Grid. In~\cite{junior2016unsupervised}, an optimum-path forest (OPF) based unsupervised NTL identification method was proposed and compared with other well-known clustering methods including \textit{k}-means and Birch. Zanetti \textit{et al.} proposed a fraud detection system (FDS) based on anomaly detection on the energy consumption reports from smart meters~\cite{zanetti2017tunable}. In their approach, an FDS state machine is designed to judge whether a grid subsystem is in an abnormal state, and unsupervised techniques of \textit{k}-means, fuzzy c-means (FCM) and self-organized map (SOM) are used to detect FDI.
	In our previous work~\cite{zheng2017electricity}, the clustering technique by fast search and find of density peaks (CFSFDP)~\cite{rodriguez2014clustering} is applied to detect load profiles with abnormal shapes. In~\cite{ahmad2017review}, various modeling techniques including supervised SVM, decision trees and bayesian networks, unsupervised OPF and real time state estimation are reviewed. Their basic assumptions, methodologies, and simulation results are presented systematically.

	In fact, the existing methods have some issues that must be addressed further. For AI-based methods, due to the difficulty in building a labeled dataset of electricity thefts, the application of classification methods is limited. Because the clustering methods are unsupervised, tampered load profiles with normal shapes can not be detected, resulting in low detection accuracy. For the state-based methods, the measurement data and system information acquisition are much more difficult to obtain. In real applications, the consumption patterns which are the focus of AI-based methods and the state consistency which is the focus of state-based methods should be both considered and utilized.
	
	In this paper, a real and general scene in which an observer meter is installed for every area containing a group of users is considered. The recorded data of the observer meter are the sum of the electricity consumptions of the area during a certain time interval. The data are available to most of the distribution system operators (DSOs) or electricity retailers. We attempt to combine the advantages of AI- and state-based methods to propose a detecting framework that adapts to the least parameters or system information to ensure general application and achieves good accuracy without any labeled training set. In particular, the maximum information coefficient (MIC)~\cite{reshef2011detecting} is used to detect the association between NTL and the tampered load profiles with minimal additional system information. Next, CFSFDP is applied to catch thieves whose load profiles are more random and arbitrary according to their abnormal density features. We ensemble the two techniques by combining the suspicion ranks to cover most types of electricity thefts. The main contributions of this paper are as follows.
	\begin{enumerate}
		\item Novel Framework: Proposing a complementary combined electricity theft detecting framework which can quantify the suspicion ranks from both the shape-similarity perspective and the magnitude-correlation perspective.
		\item New Techniques: Applying advanced and efficient machine learning techniques for abnormal detection. Specifically, MIC is used as a state-based electricity theft detecting method for correlation analysis, which only requires the observer meter data (i.e., the area total electricity consumption data) in addition to the load profiles and has high accuracy in detecting electricity thefts that appear normal in shapes; the unsupervised learning technique CFSFDP, a parameter-free method, is used to detect load profiles with unusual shapes that MIC cannot consider.
		\item Comprehensive Experiments: Conducting comprehensive numerical experiments for different types of electricity theft behaviors and comparing our method with various state-off-the-art methods to verify the effectiveness and superiority of the framework.
	\end{enumerate}
	
	The rest of this paper is organized as follows. 
	Section~\ref{sec:analysis} describes the applicable scene and gives the basic problem statement. Section~\ref{sec:framework} presents a theory of the two techniques and shows the framework of combined electricity theft detection. Numerical experiments are conducted and the evaluation results are shown in Section~\ref{sec:results}. Finally, Section~\ref{sec:conclusion} draws conclusions.
	
\begin{figure}[!t]
	\centering
	\includegraphics[width=0.4\textwidth]{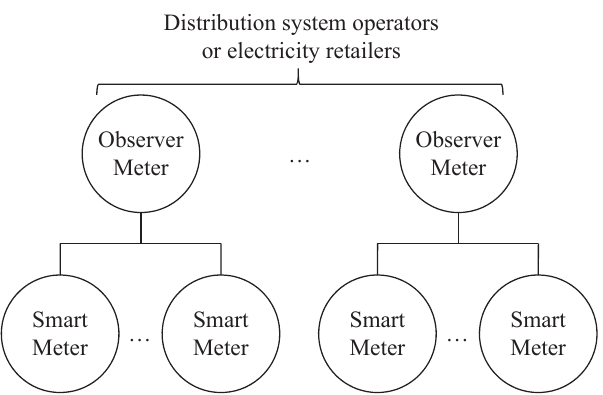}
	\caption{Observer meters for areas and smart meters for customers}
	\label{fig:observer}
\end{figure}	
	
	\section{Problem Statement}
	\label{sec:analysis}
	\subsection{Observer Meters}

	Our method is applicable to the scene of Fig.~\ref{fig:observer}, where an observer meter is installed in an area with a group of customers. An observer meter is more secure than a normal smart meter is, making it almost impossible for fraudulent users to tamper with the meter. We believe that DSOs and electricity retailers have access to the observer meter data.
	
	\subsection{False Data Injection}
	\label{subsec:FDI}
	
\begin{figure}[!t]
		\centering
		\includegraphics[width=0.43\textwidth]{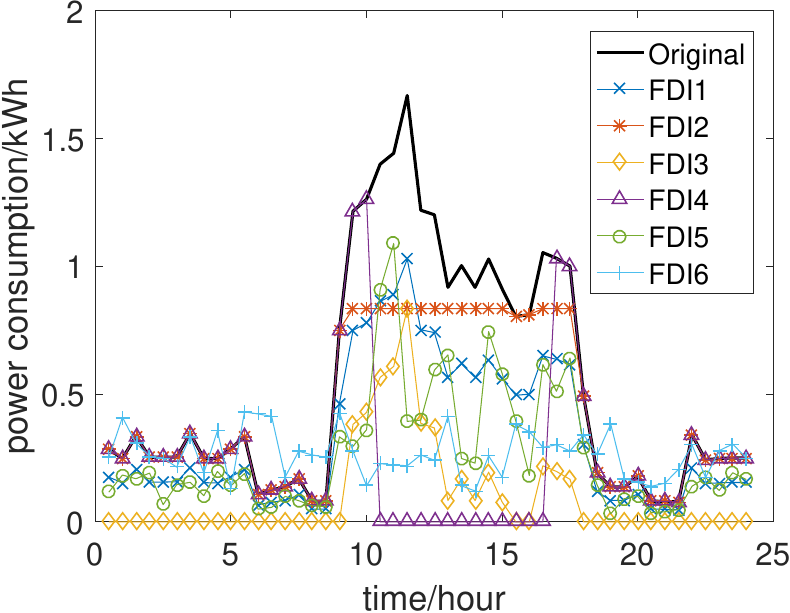}
		\caption{An example of the FDI types}
		\label{fig:FDI}
	\end{figure}

	Electricity thieves tend to reduce the quantity of their billed electricity, thus an FDI that has certain impacts on the tampered load profiles is used to simulate the tampering behaviors of the electricity thieves. We use six FDI types similar to those mentioned in~\cite{zanetti2017tunable} that have time-variant modifications on load profiles. 
	Table~\ref{tab:FDI} shows our FDI definitions, and Fig.~\ref{fig:FDI} gives an example of the tampered load profiles. In Table~\ref{tab:FDI}, $ x_t $ is the ground true power consumption during time interval $ t $, and $ \tilde{x}_t $ is the tampered data recorded by the smart meter. 
	There are many other FDI types in the literature~\cite{jokar2016electricity,han2016combating}. However, an characteristic can be generalized according to their definitions and examples: an FDI type either keeps the features and fluctuations of the original curve, or creates new patterns. This is the same for other sophisticated FDI types, so our method can handle them as well.
	\begin{table}[!t]
		\renewcommand{\arraystretch}{1.3}
		\centering	
		\begin{threeparttable}
			\caption{Six FDI types\tnote{1}~\cite{zanetti2017tunable}}
			\label{tab:FDI}
			\begin{tabular}{cl}
				\toprule
				Types & {Modification}\\
				\midrule
				FDI1 & \makecell[l]{$ \tilde{x}_t \leftarrow \alpha x_t $ \\ where $ 0.2<\alpha<0.8 $ is randomly generated}    \\
				\midrule
				FDI2 & \makecell[l]{$ \tilde{x}_t \leftarrow \begin{cases}
					x_t,\quad \text{if } x_t \leq \gamma \\
					\gamma, \quad \text{if } x_t > \gamma
					\end{cases} $ \\ 
					where $ \gamma $ is a randomly defined cut-off point, \\
					and $ \gamma <\max{\textbf{x}} $} \\
				\midrule
				FDI3 & \makecell[l]{$ \tilde{x}_t \leftarrow \max{\{x_t-\gamma,0\}} $\\
					where $ \gamma $ is a randomly defined cut-off point, \\
					and $ \gamma<\max{\textbf{x}} $} \\
				\midrule
				FDI4 & \makecell[l]{$ \tilde{x}_t \leftarrow f(t)\cdot x_t $\\
					where $ f(t) = \begin{cases}
					0,\quad \text{if } t_1<t<t_2 \\
					1,\quad \text{otherwise}
					\end{cases} $, \\ 
					$ t_1 - t_2$ is a randomly defined time period \\
					longer than 4 hours} \\
				\midrule
				FDI5 & \makecell[l]{$ \tilde{x}_t \leftarrow \alpha_t x_t $ \\ 
					where $ 0.2<\alpha_t<0.8 $ is randomly generated} \\
				\midrule
				FDI6 & \makecell[l]{$ \tilde{x}_t \leftarrow \alpha_t \bar{\textbf{x}} $\\
					where $ 0.2<\alpha_t<0.8 $ is randomly generated, \\
					$ \bar{\textbf{x}} $ is the average consumption of the load profile
				} \\
				\bottomrule
			\end{tabular}
			\begin{tablenotes}
				{\item[1] The index $ i $ in $ x_{i,t} $, $ \tilde{x}_{i,t} $ and $ \textbf{x}_i $ is omitted here for simplicity}
			\end{tablenotes}
		\end{threeparttable}
	\end{table}

	\subsection{A State-based Method of Correlation}
	\label{subsec:correlation}
\begin{figure}[!t]
		\centering
		\includegraphics[width=0.45\textwidth]{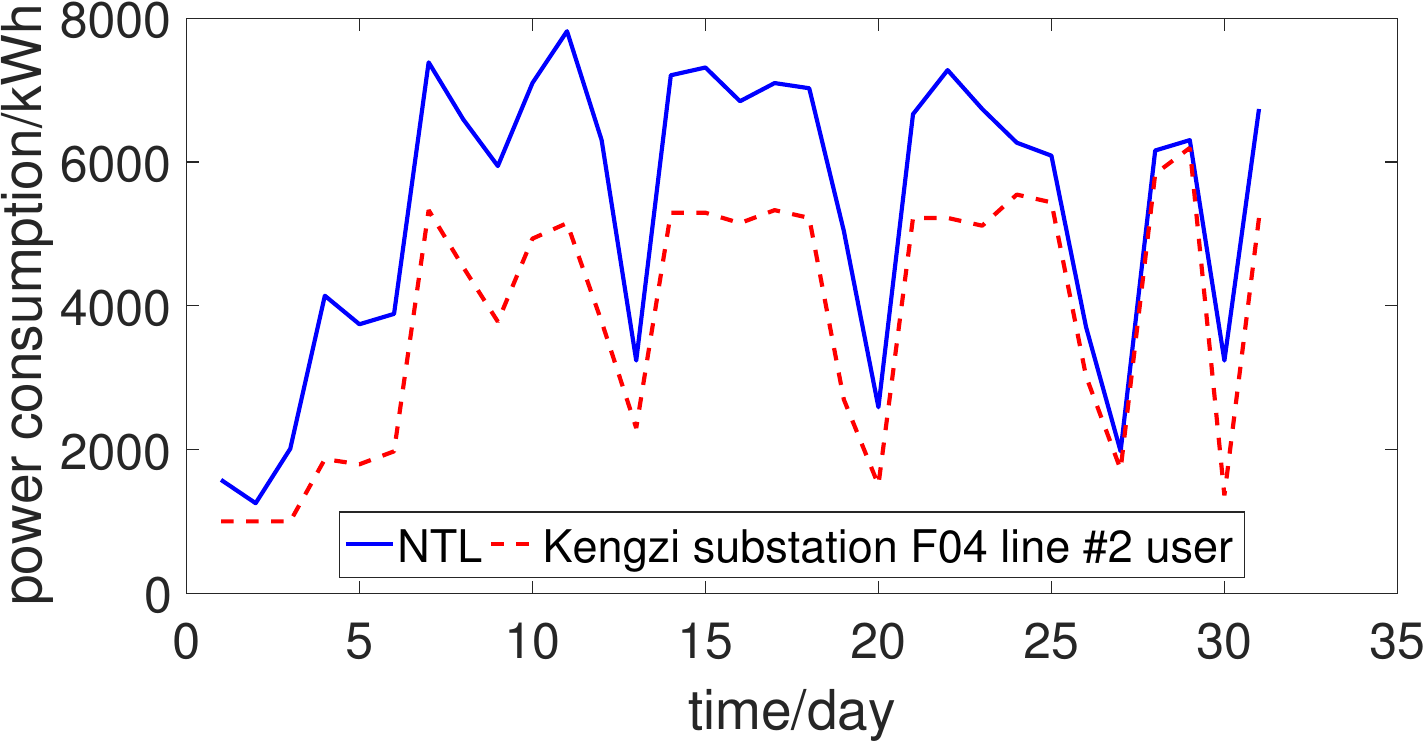}
		\caption{A real case of NTL and power consumption of the suspected user~\cite{yijia2016anomaly}}
		\label{fig:shenzhen}
	\end{figure}

	The NTL of an area $ e_t $ can be calculated by subtracting the observer meter data $ E_t $ from the sum of the smart meter data $ \tilde{x}_{i,t} $ in the area: 
	\begin{equation}
	\label{equ:NTL1}
	e_t = E_t - \sum_{i \in \mathcal{A}} \tilde{x}_{i,t}
	\end{equation}
	Let $ \mathcal{F} $ denote the set of the labels of tampered meters in the area. Eq.~(\ref{equ:NTL1}) can be represented as:
	\begin{equation}
	\label{equ:NTL2}
	e_t = \sum_{i\in \mathcal{F}} (x_{i,t} - \tilde{x}_{i,t})
	\end{equation}
	where $ x_{i,t} $ is the ground truth electricity consumption by consumer~$ i $. According to the analysis in Subsection~\ref{subsec:FDI}, if the tampered data $ \tilde{x}_{i,t} $ have a positive correlation with the ground truth data $ x_{i,t} $, then the NTL value of $ (x_{i,t}-\tilde{x}_{i,t}) $ caused by user~$ i $ is also correlated with $ \tilde{x}_{i,t} $. Because $ e_t $ is composed of several $ (x_{i,t}-\tilde{x}_{i,t}) $, the correlation between vector $ \textbf{e} $ and $ \tilde{\textbf{x}}_i $ when $ i\in\mathcal{F} $ should be stronger than the correlation when $ i\in \mathcal{B} $: 
	\begin{equation}
	\label{equ:corr}
	Corr(\textbf{e},\tilde{\textbf{x}}_i)\Big|_{i\in\mathcal{F}} > Corr(\textbf{e},\tilde{\textbf{x}}_i)\Big|_{i\in\mathcal{B}} 
	\end{equation}
	where $ Corr(\cdot,\cdot) $ is a proper correlation measurement for two vectors. 
	Fig.~\ref{fig:shenzhen} shows a real electricity theft case in Shenzhen~\cite{yijia2016anomaly}, where $ \textbf{e} $ and $ \tilde{\textbf{x}}_i $ have a high correlation. In FDI1, the correlation is linear and certain; however, in many other situations, the correlation is rather fuzzy. Note that Eq.~(\ref{equ:corr}) may not hold for some FDI types (e.g., FDI6 which produces a totally random curve); however, we can filter out a large part of electricity thefts by using Eq.~(\ref{equ:corr}). The selection of measurement $ Corr(\cdot,\cdot) $ that can precisely reveal the fuzzy relationship between NTL and tampered load profiles is of vital importance. 
	
	\section{Methodology and Detection Framework}
	\label{sec:framework}	
	
	The overall detection methodology is based on the two novel data mining techniques, i.e., MIC and CFSFDP. MIC utilize the analysis in Subsection~\ref{subsec:correlation} to detect associations between the area NTL and tampered load profiles. CFSFDP is used to determine the load profiles with abnormal shapes. According to the suspicion ranks given by the two methods, an combined rank is given to take the advantages of both methods. 
	
	\subsection{Maximum Information Coefficient}
	\label{subsec:mic}
	In statistics, the Pearson correlation coefficient (PCC) is an effective measurement for the correlation between two vectors. The PCC has a value between $ +1 $ and $ -1 $. If two vectors have a strict linear correlation, then the absolute value of PCC is 1. If two vectors are irrelevant, then the value is 0. However, the PCC cannot detect more sophisticated associations, such as quadratic or cubic, and time-variant relations. The mutual information (MI) of two variables is used as a good measurement of relevance because it detects all types of associations. MIC is based on the calculation of MI and has proved to have a better performance than MI in many occasions~\cite{reshef2011detecting}. 
	
	Given a finite set $ \mathcal{D} $ of ordered pairs, the x-values of $ \mathcal{D} $ can be partitioned into $ a $ bins and the y-values of $ \mathcal{D} $ can be partitioned into $ b $ bins. This creates an $ a $-by-$ b $ grid $ G $ in the finite 2D space. Let $ \mathcal{D}|_G $ be the distribution induced by the points in $ \mathcal{D} $ on the cells of $ G $. For $ \mathcal{D}\subset \mathbb{R}^2 $ and $ a,b\in \mathbb{N}^* $, define
	\begin{equation}
	\label{equ:I-star}
	I^*(\mathcal{D},a,b) = \max_G I(\mathcal{D}|_G)
	\end{equation}
	where the maximum is over all grids $ G $ with $ a $ columns and $ b $ rows, and $ I(\mathcal{D}|_G) $ is the MI of $ \mathcal{D}|_G $. The characteristic matrix $ \textbf{M}(\mathcal{D}) $ is defined as
	\begin{equation}
	\textbf{M}(\mathcal{D})_{a,b} = \frac{I^*(\mathcal{D},a,b)}{\log \min \{a,b\}}
	\end{equation}
	The MIC of a finite set $ \mathcal{D} $ with sample size $ |\mathcal{D}| $ and grid size less than $ B(n) $ is given by
	\begin{equation}
	MIC(\mathcal{D}) = \max_{ab<B(|\mathcal{D}|)} \{\textbf{M}(\mathcal{D})_{a,b}\}
	\end{equation}
	We use $ B(|\mathcal{D}|)=|\mathcal{D}|^{0.6} $ in this paper because it is found to work well in practice. The value of MIC falls in the range of $ [0,1] $, and a larger value indicates a stronger association.
	
	The $ MIC(\cdot) $ is applied as the $ Corr(\cdot,\cdot) $ in Eq.~(\ref{equ:corr}) to detect electricity thefts whose consumption behaviors have strong relevance to the NTL in the area.
	
	\subsection{CFSFDP-based Unsupervised Detection}
	
	To tackle the FDI types that cannot be detected by the method of correlation, we use clustering to find the outliers in the numerous load profiles. Density-based clustering methods have been widely adopted in anomaly detecting. CFSFDP~\cite{rodriguez2014clustering} is a newly-proposed method that has proved to be very powerful in large dataset clustering and outlier detection. 
	
	In CFSFDP, two values are defined for the $p$-th load profile: its local density $ \rho_p $ and its distance $ \delta_p $ from other load profiles of higher density. Both values depend on the distances $ d_{pq} $ between the data points. Eq.~(\ref{equ:rho}) gives the definition of $ \rho_p $:
	\begin{equation}
	\label{equ:rho}
	\rho_p = \sum_{q} \chi (d_{p,q} - d_c) 
	\end{equation}
	where $ d_c $ is the cut-off distance and $ \chi(\cdot) $ is the kernel function. The cut-off kernel is $ \chi(x)= \begin{cases}
	1,\text{ if } x<0\\
	0,\text{ otherwise}
	\end{cases} $. Because the local density $ \rho_p $ is discrete in Eq.~(\ref{equ:rho}), a Gaussian kernel is occasionally used to estimate $ \rho_p $, as shown in Eq.~(\ref{equ:gaussian}), to avoid conflicts:
	\begin{equation}
	\label{equ:gaussian}
	\rho_p = \sum_{q\neq p} \exp\Big[-(\frac{d_{p,q}}{d_c})^2\Big]
	\end{equation}
	The definition of $ \delta_p $ is shown in Eq.~(\ref{equ:delta}):
	\begin{equation}
	\label{equ:delta}
	\delta_p = \min_{q:\rho_q>\rho_p} d_{p,q}
	\end{equation}
	For those load profiles with the highest local density, $ \delta_p $ is conventionally written as
	\begin{equation}
	\label{equ:conventionally}
	\delta_p = \max_q d_{p,q}
	\end{equation}
	
	\begin{figure}[!t]
		\centering
		\includegraphics[width=0.38\textwidth]{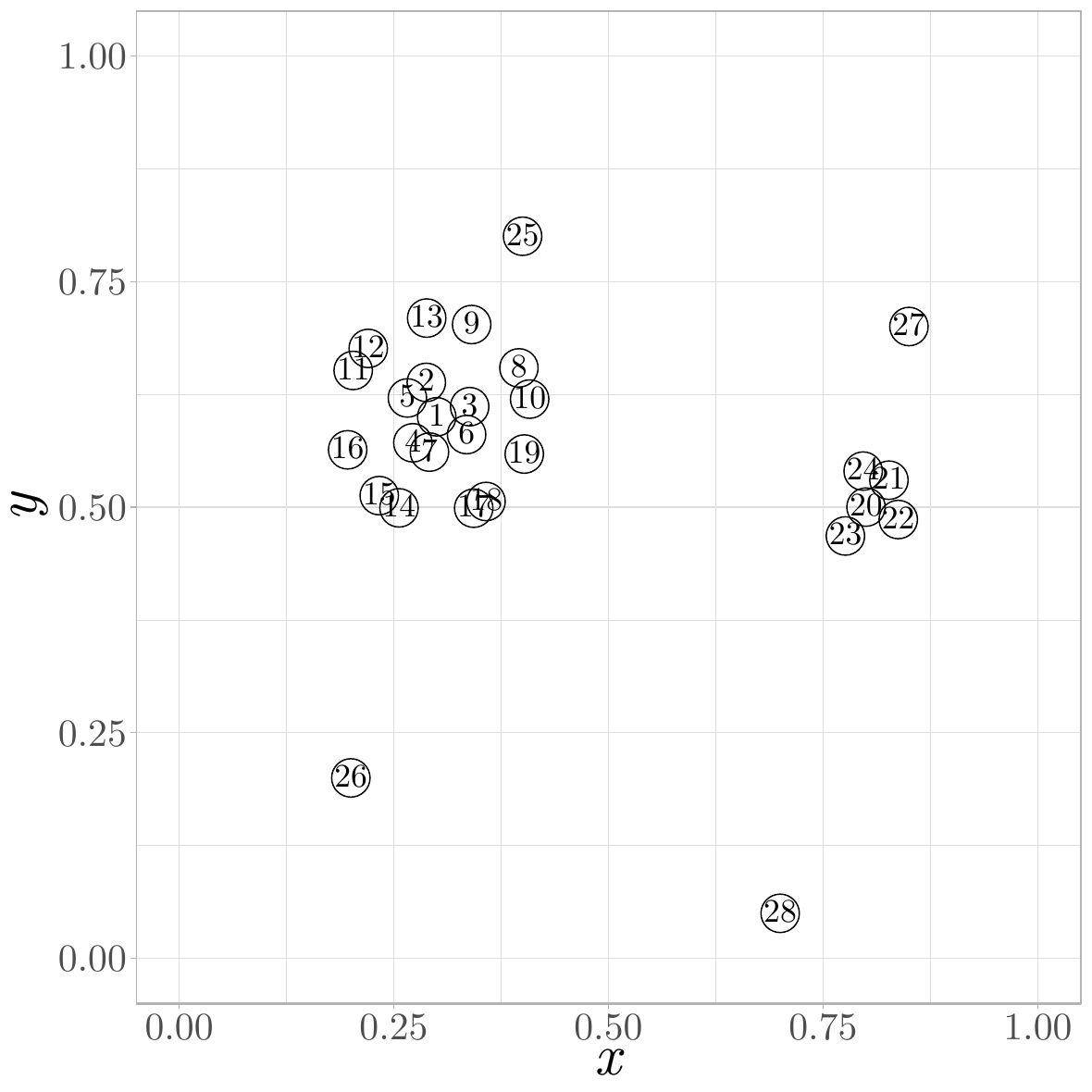}
		\caption{An example distribution of data points}
		\label{fig:data_density}
	\end{figure}
	
	\begin{figure}[!t]
		\centering
		\includegraphics[width=0.38\textwidth]{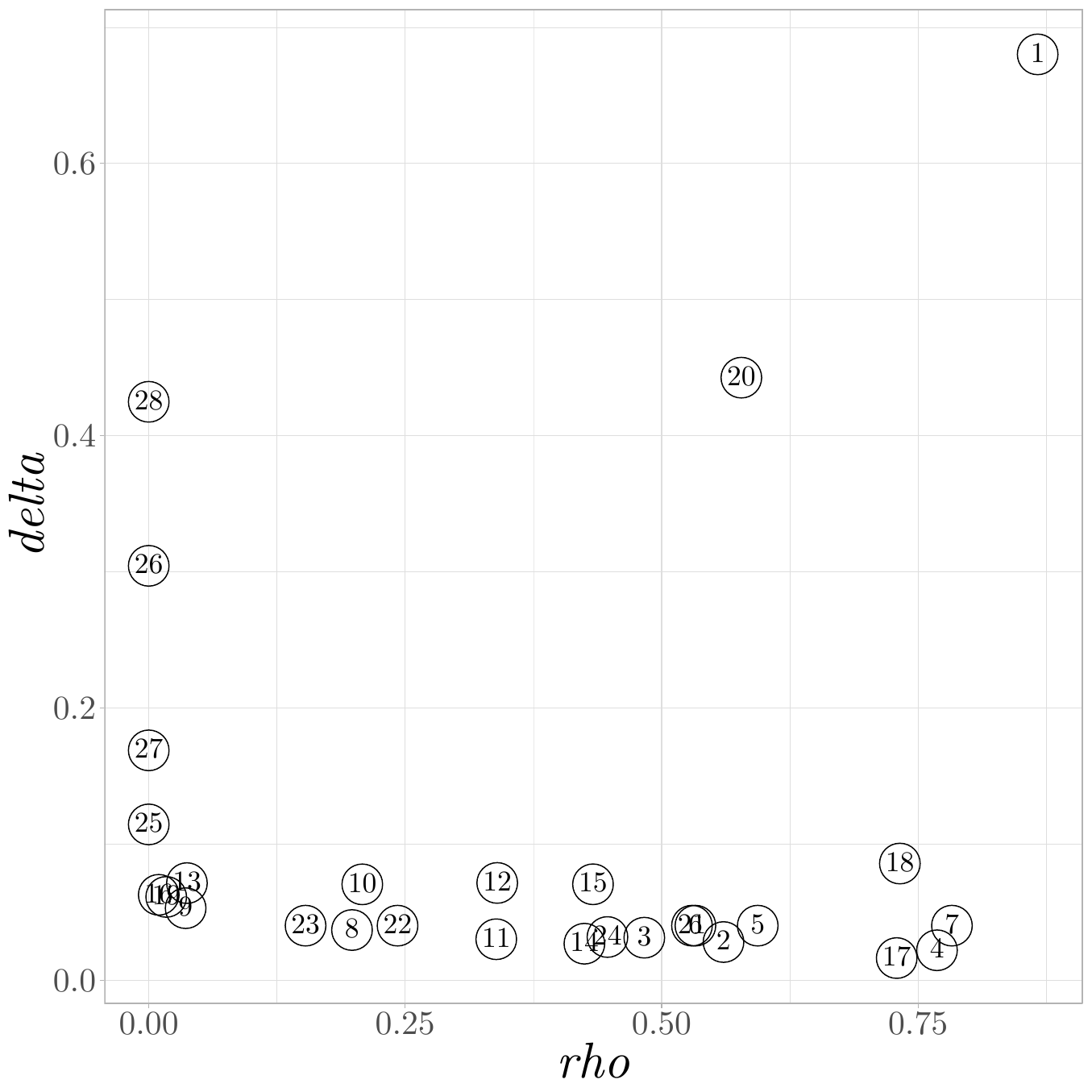}
		\caption{Scatter plot of $(\rho_p,\delta_p)$ of the example data points}
		\label{fig:decision}
	\end{figure}
	
	Although the cut-off distance $ d_c $ is exogenous in the definitions, it can be automatically chosen by a rule of thumb suggested in~\cite{rodriguez2014clustering}. 
	Fig.~\ref{fig:data_density} shows an example of 28 data points among which \#26$\sim$28 are abnormal. 
	The abnormal data points usually deviate from the normal majority, thus they only have a few neighborhood points and their distance to the high density area is larger than the normal points. From the definitions above, the spatial distribution of the abnormal points results in a small $ \rho_p $ and a large $ \delta_p $ (Fig.~\ref{fig:decision}). We define the degree of abnormality $ \zeta_p $ in Eq.~(\ref{equ:zeta}):
	\begin{equation}
	\label{equ:zeta}
	\zeta_p = \frac{\delta_p}{\rho_p + 1}
	\end{equation}
	
	Compared with \textit{k}-means and other partition-based clustering methods, density-based
	clustering can consider clusters with an arbitrary shape without any parameter selection. Moreover, the algorithm of CFSFDP is so simple that once the local density $ \rho_p $ of all the load profiles is calculated, $ \delta_p $ and $ \zeta_p $ can be easily obtained without any iteration. Load profiles with strange or arbitrary shapes are very likely to have a high value of $ \zeta_p $. Thus we can find out the abnormal load profiles according to their $ \zeta_p $ value, which is very helpful in detecting electricity thefts that MIC cannot consider. 
	
	\subsection{Combined Detecting Framework}

\begin{figure}[!t]
		\centering
		\includegraphics[width=0.3\textwidth]{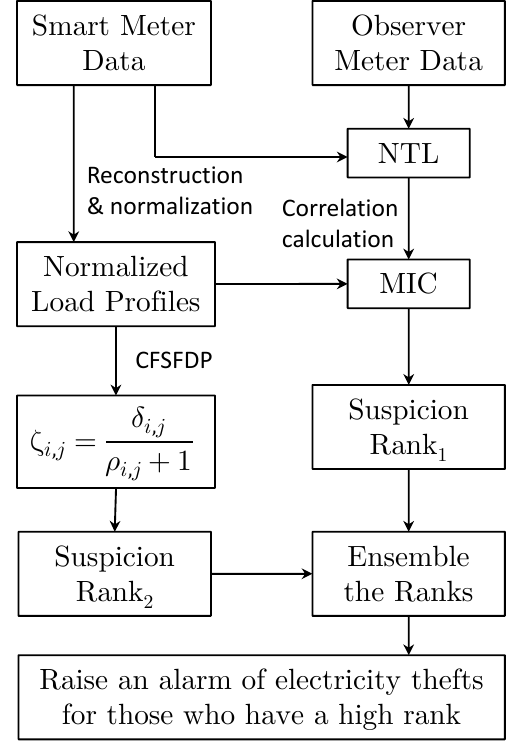}
		\caption{The detection framework of the MIC-CFSFDP combined method}
		\label{fig:framework}
	\end{figure}

	Fig.~\ref{fig:framework} shows the framework of how to utilize MIC and CFSFFDP in electricity detecting and how to combine the results of the two independent but complementary methods.
	
	For an area with $ n $ consumers and $ m $-day recorded data series, a time series of NTL is first calculated using Eq.~(\ref{equ:NTL1}). Next, we normalize each load profile $ \tilde{\textbf{x}}_{p} $ by dividing it with $ \max_t \tilde{\textbf{x}}_{p} $ and then reconstruct the smart meter dataset into a normalized load profile dataset with $ n\times m $ vectors. This procedure retains the shape of each load curve to the greatest extent and helps the clustering method focus on the detection of arbitrary load shapes. Let $ \textbf{u}_{i,j} $ denote the normalized vector of the $ i $-th consumer's load profile on the $ j $-th day and $ \textbf{e}_j $ denote the NTL loss vector of the area on the $ j $-th day. For every $ i $ and $ j $, $ MIC(\textbf{u}_{i,j},\textbf{e}_j) $ is calculated according to the equations in Subsection~\ref{subsec:mic}. Moreover, $ \rho_{i,j} $ and $ \delta_{i,j} $ are calculated using CFSFDP, and the degree of abnormality $ \zeta_{i,j} $ for vector $ \textbf{u}_{i,j} $ is obtained. 
	
	For consumer $ i $ with $ m $ MIC or $ \zeta $ values, a \textit{k}-means clustering method with $ k=2 $ is used to detect the MIC or $ \zeta $ values of suspicious days by classifying the $ m $ days into 2 groups. The mean of the MIC or $ \zeta $ values that belong to the more suspicious group is taken as the suspicion degree for consumer $ i $. Thus, the two suspicion ranks of the $ n $ consumers can be extracted by inter-comparing the $ n\times m $ MIC or $ \zeta $ values. 
	
	The idea of combining the two ranks is based on the famous Rank Product (RP) method~\cite{breitling2004rank}, which is frequently used in Biostatistics. In this paper, we use the arithmetic mean and the geometric mean of the two ranks to combine the methods, as in Eq.~(\ref{equ:rank}).
	\begin{align}
	\label{equ:rank}
	\begin{split}
	& Rank_{\text{Arith}} = \frac{Rank_1 + Rank_2}{2} \\
	\text{or }& Rank_{\text{Geo}} = \sqrt{Rank_1 \times Rank_2}
	\end{split}
	\end{align}
	Finally, a consumer is considered committing electricity theft if his combined $ Rank $ is high. 	
	
	\section{Numerical Experiments}
	\label{sec:results}
	\subsection{Dataset}
	We use the smart meter dataset from Irish CER Smart Metering Project~\cite{CER2012} that
	contains the load profiles of over 5000 Irish residential users and small \& medium-sized enterprises (SMEs) for more than 500 days. Because all users have completed the pre-trial or post-trial surveys, the original data are considered ground truth. We use the load profiles of all 391 SMEs in the dataset from July 15 to August 13, 2009. Thus, we have $ 391\times 30 = 11\,730 $ load profiles in total, and each load profile consists of 48 points, with a time interval of half an hour. The 391 SMEs are randomly and evenly divided into several areas with observer meters. For each area, several users are randomly chosen as fraudulent users, and certain types of FDI are used to tamper with their load profiles. Fifteen of the 30 load profiles of each fraudulent user are tampered with.

	\subsection{Comparisons and Evaluation Criteria}
	To demonstrate the effectiveness of our proposed method, we use other correlation analysis and unsupervised outlier detection methods for comparison: 
	\begin{itemize}
		\item Pearson correlation coefficient (PCC): a famous statistic method for bivariate correlation measurement. 
		\item Kraskov's estimator for mutual information~\cite{kraskov2004estimating}: an improved method for estimating the MI of two continuous samples. 
		\item Fuzzy C-Means (FCM): an unsupervised fuzzy clustering method. The number of cluster centers is chosen to range from 4 to 12 in this paper. 
		\item Density-based Local Outlier Factor (LOF)~\cite{breunig2000lof}: A commonly used method of density-based outlier detection. 
	\end{itemize}
	
	To obtain comprehensive evaluation results in the unbalanced dataset, we use the AUC (Area Under Curve) and MAP (Mean Average Precision) values mentioned in~\cite{zheng2018wide}. The two evaluation criteria have been widely adopted in classification tasks. The AUC is defined as the area under the receiver operating characteristic (ROC) curve, which is the trace of the false positive rate and the true positive rate. Define the set of fraudulent users $ \mathcal{F} $ as the positive class and benign users $ \mathcal{B} $ as the negative class. The suspicion $ Rank $ is in ascending order according to the suspicion degree of the users. AUC can be calculated using $ Rank $ as in Eq.~(\ref{equ:AUC}):
	\begin{equation}
	\label{equ:AUC}
	\text{AUC} = \frac{\sum_{i\in \mathcal{F}} Rank_i - \frac{1}{2}|\mathcal{F}|(|\mathcal{F}| + 1)}{|\mathcal{F}|\times |\mathcal{B}|}
	\end{equation}
	Let $ Y_k $ denote the number of electricity thieves who rank at top $ k $, and define the precision $ P\text{@}k = \frac{Y_k}{k} $. Given a certain number of $ N $, MAP@$ N $ is the mean of $ P\text{@}k $ defined in Eq.~(\ref{equ:MAP}):
	\begin{equation}
	\label{equ:MAP}
	\text{MAP@}N = \frac{\sum_{i=1}^{r} P\text{@}k_i}{r}
	\end{equation}
	where $ r $ is the number of electricity thieves who rank in the top $ N $ and $ k_i $ is the position of the $ i $-th electricity thieves. We use MAP@20 in this paper. In the random guess (RG), the true positive rate equals the false positive rate; thus, the AUC for RG is always 0.5, and the MAP for RG is $ |\mathcal{F}|/(|\mathcal{F}| + |\mathcal{B}|) $ which is the proportion of electricity thieves among all users. We consider these values to be the benchmarks. 
	
	Note that all the numerical experiments in this paper are repeated for 100 randomly generated scenarios to avoid contingency among the results. The values of AUC and MAP are calculated using the mean value to show the average performance. 
	
	\subsection{Numerical Results}
	
	In this subsection, we divide the users into 10 areas and randomly choose 5 electricity thieves for each area. Thus, each area has approximately 39 users, and the ratio of fraudulent users is 12.8\%. 
	
	\begin{figure}[htpb]
		\centering
		
		\begin{subfigure}[t]{0.86\linewidth}
			\centering\includegraphics[width=1.0\linewidth]{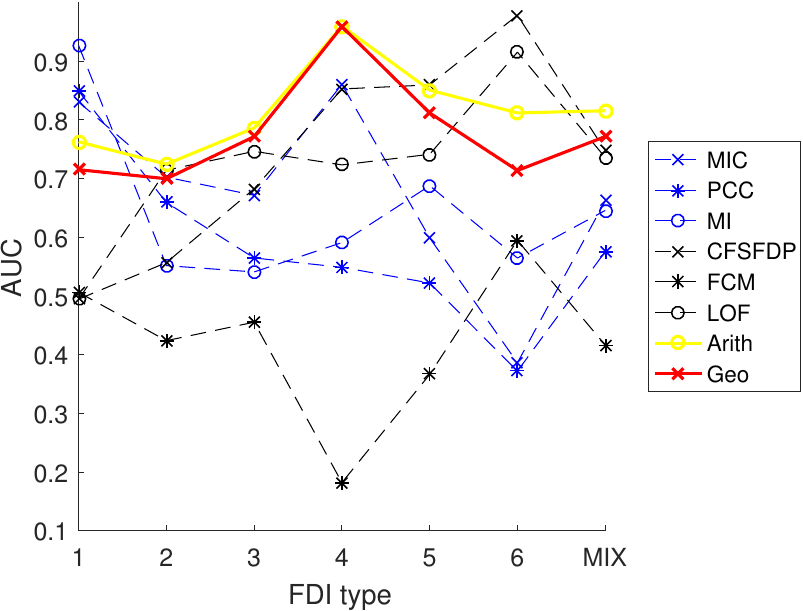}
			\caption{AUC values of the methods}
			\label{subfig:AUC_10_5}
		\end{subfigure}
		\\
		\begin{subfigure}[t]{.86\linewidth}
			\centering\includegraphics[width=1.0\linewidth]{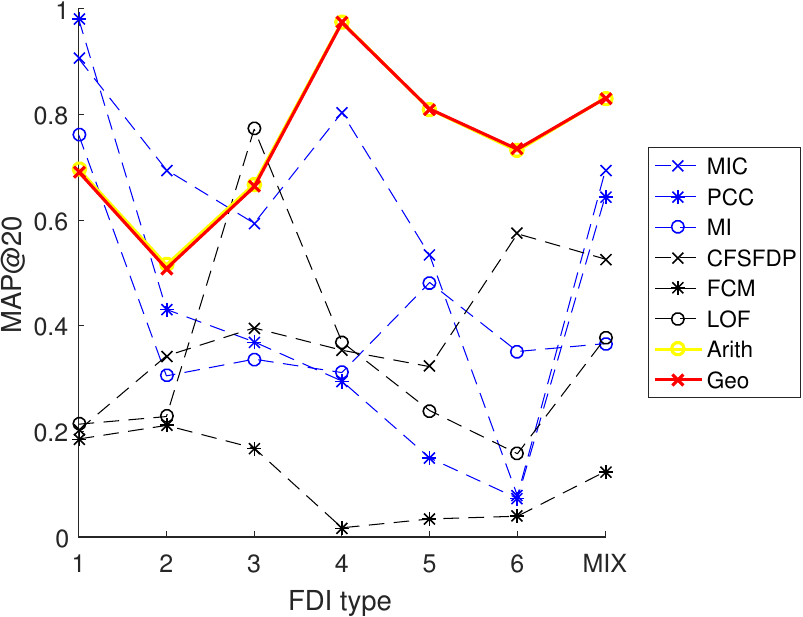}
			\caption{MAP@20 values of the methods}
			\label{subfig:MAP20_10_5}
		\end{subfigure}
		\caption{The evaluation results of the original and combined methods}
		\label{fig:MAP20_10_5}
	\end{figure}
	
	\begin{table*}[ht]
		\renewcommand{\arraystretch}{1.3}
		\caption{Average Evaluation Results of the Methods}
		\label{tab:AUC}
		\begin{center}
			\begin{tabular}{|c||c|c|c||c|c|c||c|c||c|c|c||c|c|c||c|c|}
				\hline
				\multirow{3}{*}{Type}& \multicolumn{8}{c||}{AUC(\%)}& \multicolumn{8}{c|}{MAP@20(\%)} \\
				\cline{2-17} 
				& \multicolumn{3}{c||}{Correlation} & \multicolumn{3}{c||}{Unsupervised clustering} & \multicolumn{2}{c||}{Combined} & \multicolumn{3}{c||}{Correlation} & \multicolumn{3}{c||}{Unsupervised clustering} & \multicolumn{2}{c|}{Combined} \\
				\cline{2-17}
				& MIC & PCC & MI & CFSFDP & FCM & LOF & Arith & Geo & MIC & PCC & MI & CFSFDP & FCM & LOF & Arith & Geo \\
				\hline
				\hline
				FDI1 & 83.1 & 84.9 & \textbf{92.7} & 49.5 & 50.6 & 49.5 & 76.6 & 71.5 & 90.6 & \textbf{98.1} & 76.2 & 20.2 & 18.5 & 21.4 & 69.6 & 69.1 \\
				\hline
				FDI2 & 70.3 & 66.0 & 55.2 & 55.7 & 42.4 & 71.4 & \textbf{72.5} & 70.0 & \textbf{69.5} & 43.1 & 30.5 & 34.3 & 21.2 & 22.8 & 51.5 & 50.8\\
				\hline
				FDI3 & 67.2 & 56.5 & 54.1 & 68.3 & 45.5 & 74.7 & \textbf{78.7} & 77.2 & 59.4 & 36.9 & 33.7 & 39.6 & 16.8 & 77.2 & \textbf{66.8} & 66.3 \\
				\hline 
				FDI4 & 86.1 & 55.9 & 59.1 & 85.3 & 18.2 & 72.3 & \textbf{96.0} & 95.9 & 80.4 & 29.5 & 31.1 & 35.4 & 1.7 & 37.0 & \textbf{97.5} & 97.4 \\
				\hline
				FDI5 & 59.9 & 52.2 & 68.8 & \textbf{86.0} & 36.6 & 74.1 & 85.1 & 81.2 & 53.3 & 15.0 & 48.2 & 32.3 & 3.4 & 23.9 & \textbf{81.0} & \textbf{81.0} \\
				\hline 
				FDI6 & 38.6 & 37.2 & 56.5 & \textbf{97.9} & 59.5 & 91.6 & 81.2 & 71.4 & 7.8 & 7.3 & 35.1 & 57.4 & 3.9 & 15.8 & 73.1 & \textbf{73.4} \\ 
				\hline \hline
				MIX  & \textbf{66.2} & 57.6 & 64.6 & \textbf{74.8} & 41.6 & 73.6 & \textbf{81.6} & 77.2 & \textbf{69.3} & 64.5 & 36.6 & \textbf{52.6} & 12.4 & 37.6 & \textbf{83.1} & \textbf{83.1} \\
				\hline
				
			\end{tabular}
		\end{center}
	\end{table*}
	
	Fig.~\ref{fig:MAP20_10_5} shows the comparison results of the methods. Table~\ref{tab:AUC} shows the detailed values of AUC and MAP@20 of the correlation-based methods and the unsupervised clustering-based methods for the six FDI types. The type MIX indicates that the 5 electricity thieves randomly choose one of the six types. We believe that different fraudulent users might choose different FDI types. The results for detection of single FDI types show the advantage of each method under certain situations, while the results for type MIX are of significance in practice. In CFSFDP, the cut-off kernel is used because it is faster than the Gaussian kernel and because we have a large dataset in which conflicts do not occur. In the application of FCM, there are 9 different results due to the number of cluster centers, and we only present the best among them. MI denotes the Kraskov's estimator for mutual information, and Arith and Geo are abbreviations for arithmetic and geometric mean, respectively. The best results among the 8 methods are bold for each FDI type in Table~\ref{tab:AUC}.

\begin{figure}[!t]
	\centering
	\includegraphics[width=0.43\textwidth]{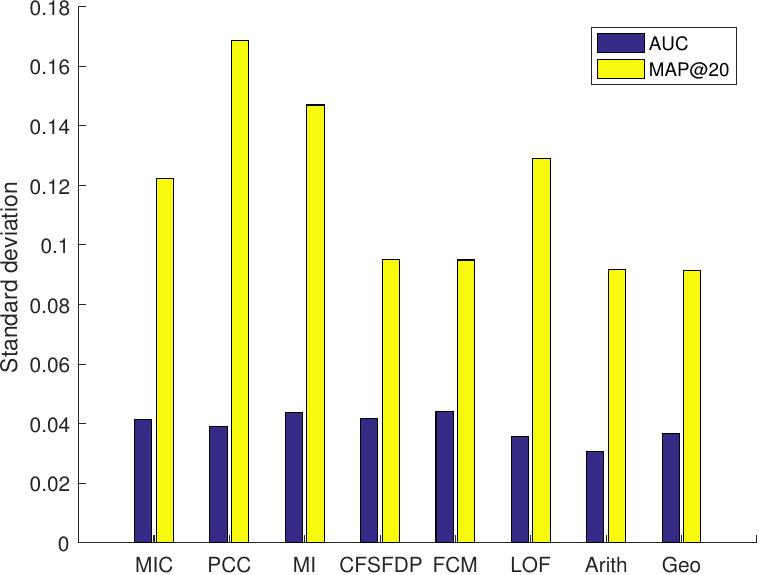}
	\caption{Standard deviations of the evaluation results}
	\label{fig:std}
\end{figure}	

\begin{figure}[!t]
	\centering
	\includegraphics[width=0.41\textwidth]{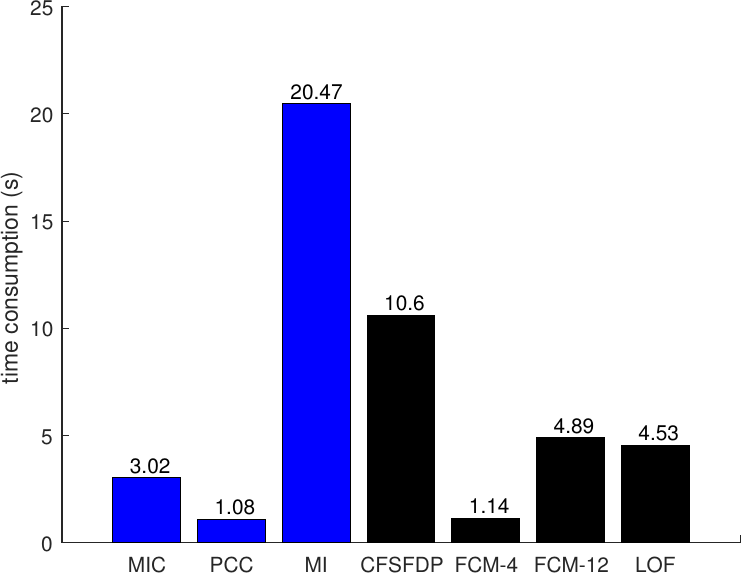}
	\caption{Time consumption of the correlation and clustering based methods}
	\label{fig:time}
\end{figure}	

	The results demonstrate that the correlation-based methods exhibit excellent performance in detecting FDI1. The blue lines in Fig.~\ref{fig:MAP20_10_5} show that MIC has a more balanced performance in both AUC and MAP@20. MIC also shows its superiority in detecting type MIX. The correlation-based method performs poorly in detecting FDI5 and FDI6 because the tampered load profiles become quite random and the correlation no longer exists. The unsupervised clustering methods, especially CFSFDP and LOF, have quite high values of AUC in detecting FDI4, FDI5, and FDI6; however, they have zero performance in FDI1 because after normalization the tampered load profiles appear exactly the same as the original load profiles. FCM have poor performance in types, except for FDI6; thus FCM may not be a good tool for electricity theft detection. Furthermore, during the numerical experiments, we noticed that the performance of FCM was heavily affected by the number of cluster centers, and it is quite unpractical in tuning the number in a wider range. From the black lines in Fig.~\ref{fig:MAP20_10_5}, CFSFDP is found to have the best performance in detecting FDI5, FDI6 and type MIX among all the clustering methods. The MAP@20 of CFSFDP is much higher than that of LOF for these types. 
	
\begin{figure}[!t]
	\centering
	\begin{subfigure}[t]{0.82\linewidth}
		\centering\includegraphics[width=1.0\linewidth]{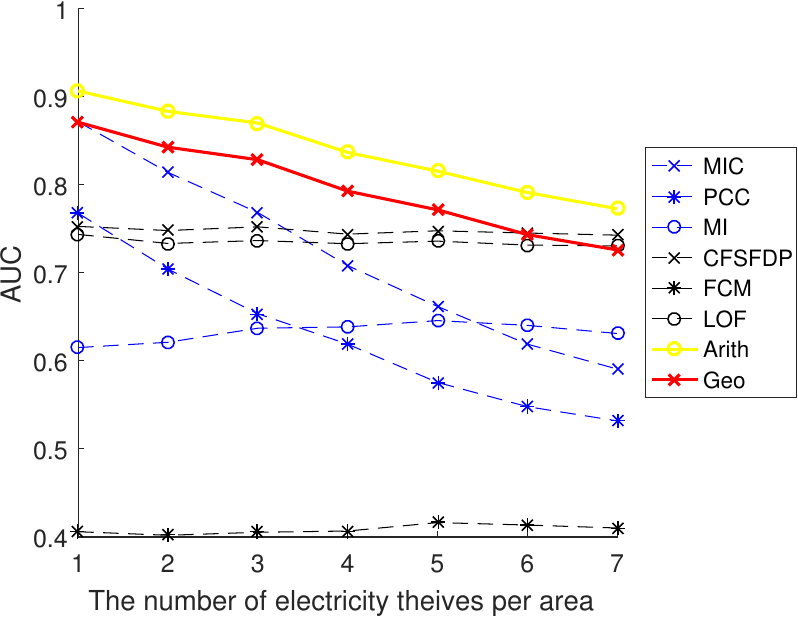}
		\caption{AUC values of the methods}
		\label{subfig:nFalse_AUC}
	\end{subfigure}
	\\
	\begin{subfigure}[t]{.82\linewidth}
		\centering\includegraphics[width=1.0\linewidth]{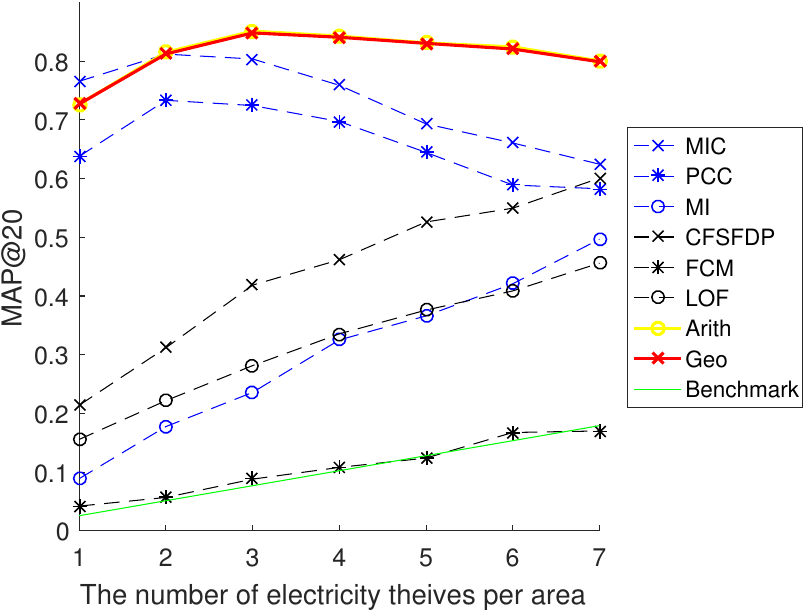}
		\caption{MAP@20 values of the methods}
		\label{subfig:nFalse_MAP20}
	\end{subfigure}
	\caption{Performance of the methods with different number of electricity thieves per area}
	\label{fig:sensi_nFalse}
\end{figure}

	The combined methods have taken the advantages of both MIC and CFSFDP. For FDI1, for which MIC specializes in, the performance of our combined methods is not as good as that of MIC. However, our methods achieves a rather high AUC of 0.766 in detecting FDI1. For FDI5 and FDI6, for which CFSFDP specializes in, our methods also have high values of AUC and MAP@20. The combined methods achieved improvements in the remaining types. The MIC-CFSFDP combined methods maintain the excellent performance of the original two methods in their own specialized situations while achieving significant improvements in the remaining situations, resulting in the best detection accuracy in type MIX and a high and steady detection accuracy for FDI1 to FDI6. The AUC value for type MIX increased from 0.748 to 0.816 (approximately 10\%), and the MAP@20 value for type MIX increasd from 0.693 to 0.831 (approximately 20\%). The results for Arith and Geo are similar in most cases, and Arith performs slightly better in AUC. 
	It is worthwhile to mention that weight factors in type MIX alter the detection accuracy. Although we assume identical weights for the FDI types, the combinded methods achieve improvements in accuracy for other non-extreme weight factors.

	Fig.~\ref{fig:std} shows the standard deviations $ \sigma $ of AUC and MAP@20 in the 100 randomly generated scenes of type MIX for each method. $ \sigma $ of AUC is approximately 4\% for all the methods, and Arith has a minimum $ \sigma_{\text{AUC}} $ of 3.08\%. $ \sigma_{\text{MAP@20}} $ is distributed between 9\% and 17\%. $ \sigma_{\text{MAP@20}} $ of Arith and Geo are 9.16\% and 9.13\%, respectively, and are smaller than that of all the other methods. The combined methods improve both the accuracy and the stability of the original methods.

	Fig.~\ref{fig:time} presents the average time consumption of the six methods for one detection of the whole $ 11\,730 $ load profiles. For FCM, we only show the results of 4 and 12 cluster centers. The test was done on an Intel Core i7-7900X@4.30GHz desktop computer with 32GB RAM. Among these methods, Kraskov's estimator for MI has the most time consumption. The combining process only requires simple calculation and sorting, and its time consumption is less than 1 s.

\begin{figure}[!t]
	\centering
	\begin{subfigure}[t]{0.82\linewidth}
		\centering\includegraphics[width=1.0\linewidth]{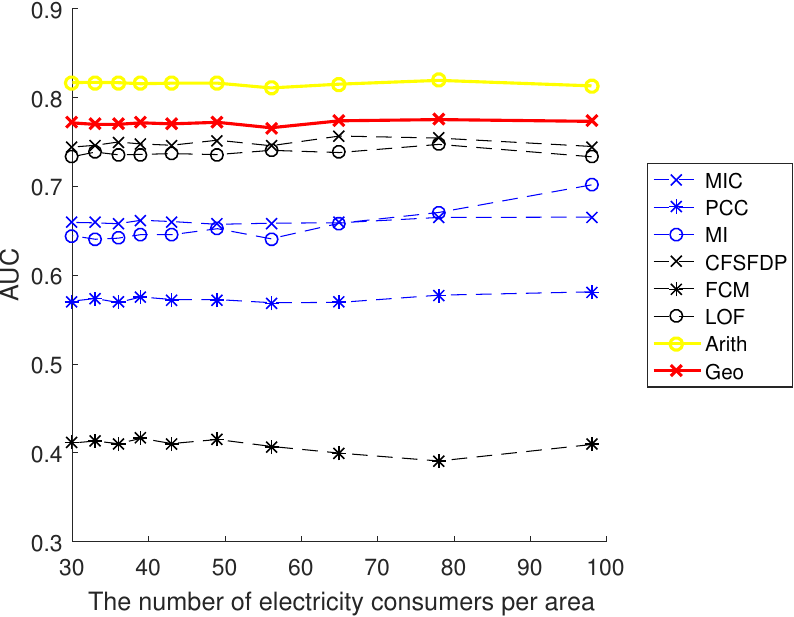}
		\caption{AUC values of the methods}
		\label{subfig:nArea_AUC}
	\end{subfigure}
	\\
	\begin{subfigure}[t]{.82\linewidth}
		\centering\includegraphics[width=1.0\linewidth]{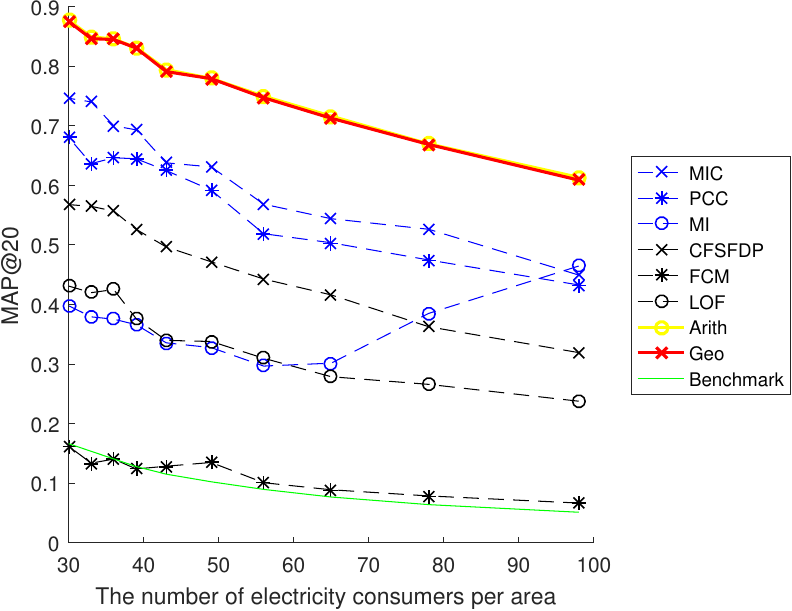}
		\caption{MAP@20 values of the methods}
		\label{subfig:nArea_MAP20}
	\end{subfigure}
	\caption{Performance of the methods with different numbers of electricity consumers per area}
	\label{fig:sensi_nArea}
\end{figure}	
	
	\subsection{Sensitivity Analysis}
	
		\begin{figure*}[!t]
			\centering
			\includegraphics[width=0.85\linewidth]{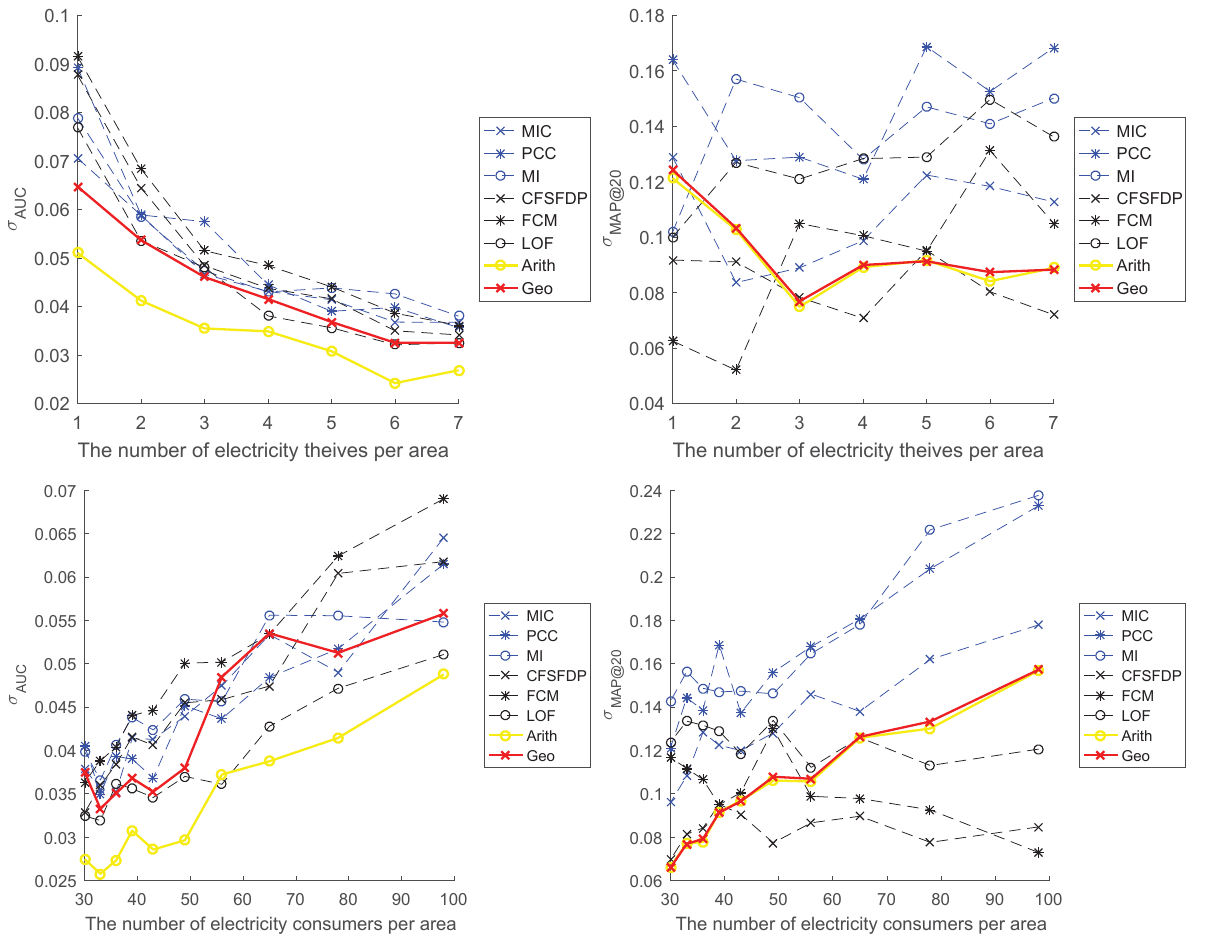}
			\caption{Standard deviations of the methods with different numbers of electricity thieves and electricity consumers per area}
			\label{fig:sensi_std}
		\end{figure*}

	When applying the electricity detection methods in real-world conditions, the number of electricity consumers or electricity thieves per area varies over a wide range, resulting in different detection accuracy and stability. In this subsection, we attempt to analyze the sensitivity in the two aspects. First, we hold the number of electricity consumers per area to 39 and change the number of electricity thieves per area from 1 to 7. Seven electricity thieves per area represent approximately 18\% of all users; this is a very severe condition. Next, we hold the number of electricity thieves per area to 5 and change the number of electricity consumers per area from 30 to 98 (which is achieved by dividing the 391 users into 4 to 13 areas). Fig.~\ref{fig:sensi_nFalse} and Fig.~\ref{fig:sensi_nArea} show the evaluation results for the two aspects of sensitivity analysis. Due to space limitations, we only present the results for type MIX.

	As the number of electricity thieves per area changes, we can see from the AUC values that MIC and PCC perform well under the conditions of fewer electricity thieves and that MI is more robust in this aspect. However, MIC and PCC perform better in MAP@20 than MI. MIC can detect electricity thieves more precisely under these conditions. CFSFDP always performs the best of the three unsupervised clustering methods. The combined method of Arith maintains excellent performance for both AUC and MAP@20.

	As the number of electricity consumers per area increases, most of the methods give a stable performance against the benchmark value. MIC is the best overall of the correlation-based methods, and CFSFDP is the best of the clustering-based methods. The combined methods achieve improvements against other methods in all conditions.

	Fig.~\ref{fig:sensi_std} shows the change in standard deviations during the two aspects of sensitivity analysis. $ \sigma_{\text{AUC}} $ shows a certain trend as the number of electricity thieves or electricity consumers increases. As the electricity theft problem becomes more severe, $ \sigma_{\text{AUC}} $ decreases slightly. The change of $ \sigma_{\text{MAP@20}} $ is more disorderly. $ \sigma_{\text{MAP@20}} $ of most methods have an upward trend as the number of electricity consumers per area increases. Although the combined methods do not always have the smallest standard deviation, the change of $ \sigma $ is over a rather small range, which is adequate for the methods in the practical application.

	\section{Conclusion}
	\label{sec:conclusion}
	This paper proposes an combined method for detecting electricity thefts against AMI in the Energy Internet. We first analyze the basic structure of the observer meters and the smart meters. Next, a correlation-based detection method using MIC is given to quantify the association between the tampered load profiles and the NTL. Considering the FDI types that have little association with the original data, an unsupervised CFSFDP-based method is proposed to detect outliers in the smart meter dataset. To improve the detection accuracy and stability, we ensemble the two techniques by combining the suspicion ranks. The numerical results show that the combined method achieves a good and steady performance for all FDI types in various conditions.

	
	%

	


	\ifCLASSOPTIONcaptionsoff
	\newpage
	\fi

	
	\bibliographystyle{IEEEtran}
	\bibliography{REF_TII-18-1861}
	
	%
	
	
	
\begin{IEEEbiography}[{\includegraphics[width=1in,height=1.25in,clip,keepaspectratio]{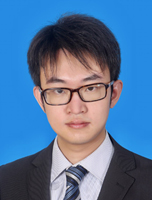}}]{Kedi Zheng}
(S'17) received the B.S. degree from the Department of Electrical Engineering in Tsinghua University, Beijing, China, in 2017. \\
He is currently pursuing Ph.D. degree in Tsinghua University. His research interests include application of machine learning and optimization algorithms in power systems.
\end{IEEEbiography}

\begin{IEEEbiography}[{\includegraphics[width=1in,height=1.25in,clip,keepaspectratio]{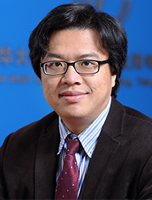}}]{Qixin Chen}
(M'10-SM'15) received the Ph.D. degree from the Department of Electrical Engineering, Tsinghua University, Beijing, China, in 2010. \\
He is currently an Associate Professor at Tsinghua University. His research interests include electricity markets, power system economics and optimization, low-carbon electricity, and power generation expansion planning.
\end{IEEEbiography}

\begin{IEEEbiography}[{\includegraphics[width=1in,height=1.25in,clip,keepaspectratio]{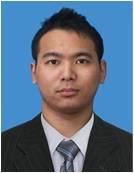}}]{Yi Wang}
(S'14) received the B.S. degree from the Department of Electrical Engineering in Huazhong University of Science and Technology (HUST), Wuhan, China, in 2014. \\
He is currently pursuing Ph.D. degree in Tsinghua University. He was a visiting student
at the University of Washington, Seattle, WA, USA from 2017 to 2018. 
His research interests include data analytics in smart grid and multiple energy systems.
\end{IEEEbiography}

\begin{IEEEbiography}[{\includegraphics[width=1in,height=1.25in,clip,keepaspectratio]{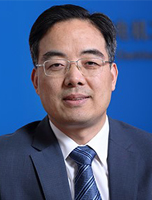}}]{Chongqing Kang}
(M'01-SM'08-F'17) received the Ph.D. degree from the Department of Electrical Engineering in Tsinghua University, Beijing, China, in 1997.\\
He is currently a Professor in Tsinghua University. His research interests include power system planning, power system operation, renewable energy, low carbon electricity technology and load forecasting.
\end{IEEEbiography}

\begin{IEEEbiography}[{\includegraphics[width=1in,height=1.25in,clip,keepaspectratio]{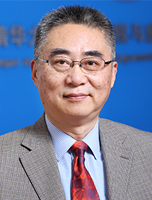}}]{Qing Xia}
(M'01–SM'08) received the Ph.D. degree from the Department of Electrical Engineering, Tsinghua University, Beijing, China, in 1989. \\
He is currently a Professor with singhua University. His research interests include electricity markets, generation scheduling optimization, and power system planning.
\end{IEEEbiography}
	
	
	\vfill
	

\end{document}